\documentclass{article}

\begin{document}

\begin{center}
{\Large{\bf Multifractals via recurrence times ?}}
\end{center}

\begin{center}
Jean-Ren{\'e} Chazottes \footnote{CPhT, CNRS-Ecole polytechnique, France, jeanrene@cpht.polytechnique.fr},
Stefano Galatolo \footnote{Dipartimento Matematica Applicata, Universit{\'a} di Pisa, Italia, galatolo@dm.unipi.it} 
\end{center}

This letter echoes the article by T.C. Halsey and M.H. Jensen published in this journal \cite{HJ}.
To say it in a nutshell, the authors evoke various methods use to determine multifractal properties
of strange attractors of dynamical systems, in particular from experimental data. They draw their
attention on a method based on recurrence times used by S. Gratrix and J.N. Elgin \cite{GE} to
estimate the pointwise dimensions of the Lorenz attractor. 

The aim of this letter is to point out that there is an important
mathematical literature establishing {\em positive} but also {\em
  negative} results about the use of recurrence times to compute multifractal characteristics of
strange attractors, e.g. \cite{BS} (to cite only one). This literature
is not cited in \cite{HJ} neither in \cite{GE}. 
In view of the use of recurrence times techniques for the experimental
demonstration of fractal properties of a variety of
natural systems, these negative results cannot overlooked. 

Recurrence times are simply defined by considering a point on the attractor and asking how long 
it takes for the orbit starting at that point to come back in a ball centered about it. Roughly
speaking, what is proven in \cite{BS} is that these return times scale like the diameter of the
ball to a power which is the pointwise dimension on the attractor.
A crucial hypothesis to prove their theorem is hyperbolicity of the
attractor. In general, recurrence times can only provide a lower bound
to pointwise dimension.

As pointed out in \cite{HJ}, the use of recurence times as a practical
tool to compute the so-called spectrum for generalized dimensions \cite{pesin}
appeared about 20 years ago and was rediscovered (and set explicitely and systematically)
in \cite{HLMV}. But the ansatz consisting in replacing the measure of
balls by the inverse return time needed to come back into them, though
working reasonably well to determine pointwise dimension, can
completely fail \cite{CU,HLMV} when used to calculate the spectrum for
generalized dimensions. The reason is essentially that return times 
and measures of balls have {\em different fluctuations at large
scales}, even in case of very "nicely" behaved dynamical systems.

More recently, hitting times instead of recurrence times
have been investigated theoretically \cite{CU,G} and turn out
to be more flexible and provide complementary informations.

In conclusion, we share the confidence of Halsey and Jensen in the
reliability of recurrence times (and hitting time as well)
in practical calculations of multifractal characteristics of attractors, but emphasize that this approach,
which emerged at a heuristic level, now relies on mathematical results showing the usefulness, but
also the {\em limitations}, of recurrence and hitting times in
capturing multifractality. Needless to say that this
field is promising both for experimentalists and mathematicians.

\end{document}